\documentstyle[preprint,aps,amssymb]{revtex}
\tightenlines 

\newcommand{\ket}[1]{\left| #1\right>}
\def\al{\alpha}

\def\de{\delta}
\def\eps{\epsilon}
\def\si{\sigma}
\def\t{\tau}

\def\NPB{Nucl. Phys. }
\def\PLB{ Phys. Lett.  }  

\def\PRD{Phys. Rev.  }

\def\JHEP{ J. High Energy Phys. }
\begin{document}
\preprint{\vbox{\noindent
\null\hfill INFNCA-TH00017}}
%
%
% \draft command makes pacs numbers print
\draft
\title{Open Strings, Holography and  Stochastic Processes}
\author{Mariano Cadoni\footnote{E--Mail: cadoni@ca.infn.it} and Paolo
Carta\footnote{E--Mail: carta@ca.infn.it}} \address{Universit\`a degli
Studi di Cagliari, Dipartimento di Fisica and\\INFN, Sezione di
Cagliari, Cittadella Universitaria 09042, Monserrato, Italy.}
\maketitle
\begin{abstract}
We use the correspondence between string states and local operators on
the world-sheet boundary defined by vertex operators in open string
theory to put in correspondence, holographically, the bosonic open
string with the large $N$ limit of a mechanical system living
on the world-sheet boundary.  We give a natural interpretation of this 
system in terms of a one-dimensional stochastic
process and show that the correspondence takes the form of a map
between two conformal field theories with central charge $c=24$ and
$c=1$.
\end{abstract}
\pacs{} .
\section{Introduction}
Hundred years ago the German physicist Max Plank solved the problem of
black body radiation, using an idea that can be reinterpreted in view
of the recently proposed holographic principle. He started with a 
three-dimensional (3D)
ensemble of harmonic oscillators affected by a trivial ultraviolet
divergence. Then, introducing a fundamental scale, he regularized the
theory and ended up with a four-dimensional (4D) statistical field theory.

Recent exiting developments in high energy physics, in particular the
holographic principle and the Anti--de Sitter/Conformal Field Theory
(AdS/CFT) correspondence \cite{adscft}, suggest that a lower
dimensional version of the Planck mechanism may be at work also in
this context. In fact,  it has been shown that conformal field theories
(CFT) in $d=2$ can be put in correspondence with the large $N$ limit of
 a mechanical system. More in detail, investigations of the AdS/CFT
correspondence in two dimensions \cite{adsdue} show that 
two-dimensional (2D) gravity
can be described in terms of an ensemble of simple mechanical systems
(representing a De-Alfaro-Fubini-Furlan (DFF) \cite{dff} model) \cite{CCKM}.
Another piece of evidence in this direction is the large $N$ limit of
Calogero models, which can be shown to describe a CFT in $d=2$
\cite{CCK} (The large $N$ limit we consider in this paper should not
be confused with the usual thermodynamic limit, where the particle
density is kept constant).

One may therefore conjecture that the Plank idea represents the
simplest realization of a more general principle, which has
holographic nature and relates the large $N$ limit of mechanical
systems with field theories (in particular CFT and gravity) in
$d\ge2$.

In this paper we test this conjecture in a simple context, namely open
strings with 2D target space.  The vertex operators in string theory
define a natural, holographic, correspondence between string states
and local operators on the world-sheet boundary. We use this
correspondence to relate the open string theory with the large $N$
limit of a mechanical system living on the world-sheet boundary.

Since it is known that Wiener processes may be realized in terms of an
ensemble (or if you want a large $N$ limit) of independent normal
modes \cite{ito}, it turns out that the most natural framework to
describe our correspondence is to consider stochastic processes in
$d=1$.  We will show that the correspondence takes the form of a map
between two CFT's. Moreover, the mechanical system has a natural
interpretation in terms of an ensemble of decoupled harmonic
oscillators. To regularize the short distance behavior of the system
we have to resume the hundred-years old Plank regularization
procedure.

The structure of the paper is as follows. In sect. II we 
interpret the
vertex operators in open string theory as a holographic
correspondence. In sect. III we use them to construct the physical
spectrum of open string theory in $D=2$. In sect. IV we consider a
one-dimensional (1D) stochastic process and show that it can be put in 
correspondence,
both at level of spectrum and Hilbert space, with open string theory.
Finally in sect. V we state our conclusions.

\section {Open strings, vertex operators  and the holographic principle}
In string theory there is nice correspondence between string states 
and local operators (see for instance Ref. \cite{pol}).  This
correspondence is realized explicitly using the vertex operators
associated with the string states. Although this fact is a general
feature of string theory, for open strings it becomes a genuine
realization of the holographic principle in two spacetime dimensions.
In the case of open strings the world-sheet, whose coordinates are
$\tau$ and $\sigma$, has a timelike boundary (the line swept out by the
strings endpoints at $\sigma=0,\pi$) so that the correspondence takes
the form of an isomorphism between string states on the
two-dimensional ``bulk'' and local operators on the one-dimensional
boundary.

In the following will focus on the bosonic open string with 2D target
space. Since the above correspondence does not depend on the
target space dimension, our idea is expected to work in any
dimension.
  We impose Neumann boundary conditions ($n^{a}\partial_{a}
X^{\mu}(\tau,\sigma)|_{\sigma=0,\pi}=0$).  The string field $X^\mu$ has the
normal mode expansion (we will follow the conventions of
Ref.\cite{GSW})
\begin{equation}\label{e:mode}
X^\mu (\t,\si) = x^\mu +p^\mu \tau
+i\sum_{n\ne0}\frac{\al^\mu_n}{n}e^{-in\t}\cos(n\si),
\end{equation}
where $\mu=0,1$, $-\infty\le\t\le \infty$ and $0 \le\si \le
\pi$. 

We will show that, although very simple, the case of a 2D target space
is physically rich enough to give a non trivial realization of the
holographic principle. Moreover, it has been shown that the
AdS/CFT correspondence in $d=2$ can be realized in terms of a 2D
bosonic string with Neumann or Dirichlet boundary conditions
\cite{CC}. Therefore, the string-states/local-operators isomorphism
discussed here can be also considered as an explicit realization of
the AdS$_{2}$/CFT$_{1}$ correspondence.

The explicit form of the string states/local-operators correspondence can be
easily worked out. Passing to complex world-sheet coordinates $z,\bar
z$ and expressing $\al^\mu_n$ as contour integral in the complex plane
one gets \cite{pol}
$$
\al^\mu_{-m}= {2i\over (m-1)!} \partial^{m}X^\mu (0,0).
$$
This equation defines a correspondence between  string states and
(normal-ordered) local operators $\cal A$ at the origin of the complex
plane.  For instance we have for the vertex operators,
\begin{equation}\label{e2}
\ket{0,p}\cong : e^{i p_{\mu}X^\mu (0,0)}: .
\end{equation}
Considering time evolution, generated by the string Hamiltonian $H$,
$$
{\cal A}(\tau)= e^{i\tau H} {\cal A}(0,0) e^{-i\tau H},
$$
the correspondence holds for local operators $\cal A(\tau)$, so that 
we are dealing with an holographic correspondence between 2D bulk 
string states and  local operators evaluated on the 1D timelike 
boundary $\sigma=0,\pi$.
If ${\cal A}(\tau)$ has conformal dimension $J$ we have, 
\begin{equation}\label{e4}
[L_m,{\cal A}(\tau)] =e^{im\tau}\left(
-i\frac{d}{d\t}+mJ\right) {\cal A}(\tau),
\end{equation}
where $L_{m}$ are the Virasoro generators.  When applied to the vertex
operator $\exp (i p_{\mu}X^\mu )$ of Eq. (\ref{e2}), the previous
equation gives a relation between conformal dimension of the operator
and mass $M$ of the string state\cite{GSW}: $J=-M^{2}/2$. This
equation represents the two-dimensional version of the general
relation, found in the context of the AdS/CFT correspondence, between
dimensions of operators of the conformal field theory and masses of
states of AdS-gravity \cite{witten}.

In the next sections we will explore the physical content of the 
holographic correspondence between 2D open string states and local
operators on the world-sheet boundary.

\section{Vertex operators and the string spectrum }
At first sight the 2D bosonic string theory seems rather trivial:
because there are no transverse directions, all the local degrees of
freedom can be gauged away.  A closer investigation reveals a
topological obstruction that prevents this.  For a $2D$ target space 
it is impossible to completely
fix the reparametrization and Weyl scaling gauge symmetry and to
impose Neumann boundary conditions avoiding at the same time that the
2D world-sheet degenerates into a line.  On the other hand, it is
evident that in case of a 2D target space we cannot resort to a 
non covariant formalism as in the $D>2$ case (light-cone
quantization for instance). In view of the holographic correspondence 
we are going to discuss, one may argue that the bulk
gauge degrees of freedom become in some sense physical on the
boundary.

In this section we aim to construct the Hilbert space for the bosonic 
open string with 2D target space.
 This can be easily done by considering it  as a particular case of the general
formulation of the bosonic string (see e.g. Ref. \cite{GSW}). It can
be shown that the Hilbert space has a fairly simple structure: it is
the Verma module of an infinite dimensional conformal algebra with
central charge $c=24$.

In order to fix the notations let us briefly review a few basics about
string quantization. The normal modes $\al^{\mu}_n$ appearing in
Eq. (\ref{e:mode}) are interpreted as operators obeying the usual
quantization conditions:
$[\al^\mu_n,\al^\nu_m]=n\de_{m+n}\eta^{\mu\nu}$ and
$[x^\mu,p^\nu]=i\eta^{\mu\nu}$. We will use light-cone coordinates in the
target space: $\eta_{+-}=\eta_{-+}=-1$.  The Hilbert space is spanned
by the base of states given by
\begin{equation}\label{hilbert}
\prod_{\mu=0,1}\prod_{n=1}^\infty (\al_n^\mu)^{\eps_{n,\mu}}\ket{0,p},
\end{equation}
where $\eps_{n,\mu}=0,1,\ldots$ are the occupation numbers. The
physical states are defined by the Virasoro conditions
\begin{equation}\label{e:vir}
(L_m -a\delta_m)\ket{\phi}=0 \qquad m\ge 0, 
\end{equation}
where $a$ is a $c$ number, the conformal weight of the vacuum. We have
to eliminate the unphysical ghosts (negative norm states) imposing the
constraints (\ref{e:vir}) in a way consistent with the Lorentz
invariance of the theory.  In $D=2$ we have the peculiarity that there
are no effective normal modes. Indeed, it is well known \cite{br} that
the spectrum generating algebra for the 2D system is isomorphic 
to the conformal algebra 
generated by $L_m$.  The physical spectrum can be found using the Brower
construction \cite{br}, a version of the old--fashioned covariant
formalism of Del Giudice, Di Vecchia and Fubini \cite{DDF}.

In this approach a set of operators that commute with the generators
$L_m$ are constructed. These operators acting on the ground state
$\ket{0,p}$, give the physical states of the theory.  It is important
to choose properly the vacuum, that is to fix the highest weight 
state,
$L_0\ket{a} = a\ket{a}$. On general grounds we know that in order to
obtain a ghost free spectrum, we must impose the condition $a \le1$.

The mass-shell condition fixes the ground state:
\begin{eqnarray*}
&&L_0 = \alpha' p^2 + \sum_{n>0}\alpha_{-n}\cdot\alpha_{n} \nonumber\\
&&(L_0-a) \ket{a} =0 \Rightarrow \alpha' p^2 =a.
\end{eqnarray*}
In the following we will set $\al' =1/2$ and choose $p^+ = 1$ and
$p^{-}=-a$.  Note that the ground state is tachionic only for $0<a\le
1$.  If we define $k_n$ such that $k_n^+=0$, $k_n^-=-n$, then the
vertex operators defined on the world-sheet boundary, $V(k_n,\tau)=\exp(ik_n\cdot
X(\tau,0))$, are periodic functions of $\t$ with period $2\pi$. We
consider the vertex operators \cite{br}
\begin{equation}\label{e:Vbr}
V^-(k_n,\t) = : \dot{X}^-e^{inX^+}: -\frac{1}{2}in\frac{d}{d\t}(\log
\dot{X}^+) e^{inX^+}.
\end{equation}
It is straightforward to check using Eq. (\ref{e4}) that  $V^-(k_n,\t)$ 
has conformal
dimension $J=1$.  The second term in
Eq. (\ref{e:Vbr}) is exactly what is needed to cancel the anomalous,
$k_n$ dependent, dimension due to the normal ordering in the first
term. It follows that the operators
$$
A^-_n = \frac{1}{2\pi}\int_0^{2\pi}d\t \,V^-(k_n,\t)
$$
trivially commute with the Virasoro generators: $[L_m,A^-_n]=0$ and can
be used to define the physical states. With the redefinition $\tilde
A^-_n = A^-_n + \delta_n$ we find the commutation relations:
\begin{equation}\label{e:alg}
[\tilde A^-_n, \tilde A^-_m] = (n-m)\tilde A^-_{n+m} +2(n^3-n)\delta_{n+m}.
\end{equation}
This is a Virasoro algebra with central charge $c=24$. It is easy to
check that 
$$
\tilde A^-_n \ket{0,p} =0 \qquad n>0.
$$
Moreover, since $\tilde A^-_0 \ket{0,p} = (p^-+1)\ket{0,p} =
(1-a)\ket{0,p}$, we find that $\ket{0,p}$ is a highest weight
state for the algebra (\ref{e:alg}) with weight $h=1-a$. The case
$a=1$ will be excluded since we want to avoid a zero norm ground
state. 

It is a corollary of the no-ghost theorem in $D$ dimensions that
physical states in our case are given by the Verma module $V(c,h)$,
representing the physical states of the full Hilbert space
(\ref{hilbert}), with $c=24$ and $h=1-a$,
\begin{equation}\label{e:Verma}
V(24,1-a) = \{\tilde A^-_{-n_1}\tilde A^-_{-n_2}\ldots \tilde
A^-_{-n_k}\ket{0,p} \, \vert \, 1\le n_1\le \ldots \le n_k; k >0 \}.
\end{equation}
All the states (\ref{e:Verma}) have the same energy: $L_0 \ket{\phi}=a
\ket{\phi}$, $\forall \ket{\phi} \in V(24,1-a)$, and are annihilated
by the conformal generators $L_m$ for $m>0$. It is important to
observe that the module $V(24,1-a)$ does not contain any null
submodule, so that it gives an irreducible representation of the
Virasoro algebra: $V(24,1-a)=M(24,1-a)$, $M(c,h)$ being the
irreducible representation obtained quozienting out of the Verma
module $V(c,h)$ the null submodules. Indeed, the Kac determinant never
vanishes in the region $c>1$ and $h>0$ (see for instance
Ref. \cite{dif}). This means that all the states at a given level of
the Verma module $V(24,1-a)$ are linearly independent.

\section{One-dimensional stochastic processes and holography}
In the previous section we have identified precisely the physical
Hilbert space of the 2D bosonic string.  In this section we will try
to realize the holographic idea.  We will look for a mechanical
system, defined on the timelike boundary of the open string
world-sheet, whose Hilbert space can be mapped exactly on the
physical Hilbert spaces of the string (\ref{e:Verma}).

Trying to setup this correspondence one can hardly escape a basic
problem: how can a one-dimensional system have the same degeneracy of
a 2D field theory? To be more concrete let us consider the simple example
of DFF conformal quantum mechanics \cite{dff}. Since the compact
operator $R$, basically the Hamiltonian, of Ref. \cite{dff} has the
 same spectrum of the harmonic
oscillator, $r_n = r_0 +n$, where $r_0$ is a constant depending on the
coupling and $n \ge 0$, neglecting the zero point term the free
energy is given by $F=k_BT \ln(1-e^{-\beta})$, where $\beta = 1/
k_BT$. On the other hand, for a  2D field theory we expect a
Stefan--Boltzmann behavior $F\propto T^2$ and  an entropy $S \propto
T$. This fact indicates that in $d=1+0$, an interaction between a
finite number of particles cannot give the required degeneracy. A
possibility left open is to consider the limit $N\to\infty$ of a
system of $N$ interacting particles, whose interaction is possibly
described by a Calogero model (see for instance Ref. \cite{gb}).

In this paper
we explore a situation similar to that described in
Ref. \cite{CCKM}. In that paper it was shown that the quadratic
dependence of the free-energy from the temperature may emerge from the
coupling of the DFF model with an arbitrary time-dependent external
source.  Along this line of thoughts it seems natural to consider
stochastic processes.

Our starting point is a simple idea based on a well-known property of
the Wiener process: if $\al_0,\al_1,\ldots$ are a sequence of Gaussian
variables, each with the same distribution $(2\pi)^{-1/2}\exp
-\al^2/2$, then the coordinate $y(t)$ of the stochastic process in
$d=1$ may be written as
\begin{equation}\label{wie}
y(t) = \frac{t}{\sqrt{\pi}}\al_0 + \left(
\frac{2}{\pi}\right)^{\frac{1}{2}}\sum_{n>0} \frac{\sin{nt}}{n}\al_n.
\end{equation} 
Indeed a Gaussian process is completely determined by its covariance
and the covariance of the random function $y(t)$, inferred from the
distribution of the $\al$, is the same as the covariance of a Wiener
process for $0\le t\le \pi$, with sample paths in one dimension and
diffusion coefficient equal to $1/2$ \cite{ito}. Upon quantization, on
the basis of the spectrum given by the $\al$ modes it is conceivable
to extract, under certain conditions, an infinite dimensional
conformal algebra, or what is the same, to define a mapping between the
Hilbert space of this system and the one of the 2D open string
theory described in the previous section.

It is worth observing that although (\ref{wie}) has a close
resemblance with the open string mode expansion (\ref{e:mode}) 
evaluated on the
$\sigma = 0$ boundary, the correspondence between the 1D and 2D
theories is not immediate. In particular the normal modes of the
string in Eq.  (\ref{e:mode}) cannot be identified with the modes
$\al$ appearing in Eq. (\ref{wie}).  The crucial point is that in the
string case we have the Lorentz (causal) structure, typical of any
local field theory, which induces, upon quantization, negative norm
states. In the physical theory arising after the eliminations of the
ghosts from the spectrum, there are no more normal modes but only
Virasoro operators generating a Verma module.  On the other hand the
1D stochastic process does not know anything about Lorentz invariance,
so that the normal modes $\al$ appearing in Eq. (\ref{wie}) generate
directly the physical spectrum.  For this reason the correspondence
between the two theories can be made only at the level of the
associated physical spectra and physical Hilbert spaces.

Basically to each path $y(t)$, defined by Eq. (\ref{wie}) in a finite
time interval $[0,L]$ we associate an energy
\begin{equation}\label{e:base}
E[y] = \frac{1}{2}\int_0^L dt\, \left(\frac{dy}{dt}\right)^2.
\end{equation}
Physically $E[y]$ may be interpreted as the energy dissipated in a
thermal reservoir by a random perturbation with a finite temporal
extension. Of course the most natural interpretation would be to think
of $E$ as the energy loss by a Brownian particle driven by some
external force, but in this case the temporal boundary conditions
should be chosen differently.  If we consider $t$ as a space
dimension, instead of a temporal one, a simple mechanical
interpretation is to think of $E$ as the elastic energy of a
weightless elastic string clamped at the endpoints \cite{kam}.

In a statistical treatment to each path is associated a probability
factor
\begin{equation}\label{e:libero}
e^{-\beta E} = \exp\left\{ -\frac{1}{2}\beta \int_0^L dt\,
\left(\frac{dy}{dt}\right)^2\right\}.
\end{equation}
Discretizing the energy integral it is easy to define a stochastic
process $Y(t)$ whose sample functions are the paths $y(t)$
\cite{kam}. The stochastic process may be described as follows. If we
choose $n$ different points in the interval $(0,L)$ and label them
with increasing time
\begin{equation}\label{e:ord}
0< t_1 < t_2 < \ldots <t_n < L, 
\end{equation}
the energy of a piece-wise straight path is given by
$$
\frac{1}{2}\sum_{k=0}^n\frac{(y_{n+1}-y_{n})^2}{t_{n+1}-t_{n}},
$$
where $t_0=0$, $t_{n+1}=L$, $y_0=y_{n+1}=0$. A stochastic process
$Y(t)$ is completely specified once is defined a hierarchy of
functions $P_n(y_1,t_1;y_2,t_2;\ldots;y_n,t_n)$, $n \in {\Bbb N}$,
giving the joint probability densities that $Y$ has the value $y_1$ at
the time $t_1$, $y_2$ at the time $t_2$ and so on. From  $P_n$ we
compute the correlations:
$$
\langle Y(t_1)Y(t_2)\ldots Y(t_n)\rangle= \int y_1y_2\ldots y_n\,
P_n(y_1,t_1;y_2,t_2;\ldots;y_n,t_n)\, dy_1dy_2\ldots dy_n .
$$
In order to specify a stochastic process the hierarchy of probability
distributions has to obey the four consistency conditions (Kolmogorov)
\begin{eqnarray}\label{e:Kolmo}
(i)&& \,P_n \ge 0; \nonumber \\ (ii)&& \, P_n \,\,\text{is a symmetric
function of the pairs}\,(y_k,t_k); \nonumber \\ (iii)&& \, \int
P_n(y_1,t_1;y_2,t_2;\ldots;y_n,t_n)\, dy_n =
P_{n-1}(y_1,t_1;y_2,t_2;\ldots;y_{n-1},t_{n-1})\nonumber \\
(iv)&& \, \int P_1(y_1,t_1)\, dy_1 =1.
\end{eqnarray}
Given the time-ordering (\ref{e:ord}), we define the hierarchy
satisfying the conditions (\ref{e:Kolmo}) by setting,
\begin{equation}\label{e:proc}
P_n(y_1,t_1;y_2,t_2;\ldots;y_n,t_n) = \left( \frac{2\pi
L}{\beta}\right)^{1/2}\prod_{k=0}^n \left(
\frac{\beta}{2\pi(t_{k+1}-t_k)}\right)^{1/2}\exp \left[
-\frac{\beta}{2} \frac{(y_{n+1}-y_{n})^2}{t_{n+1}-t_{n}}\right], 
\end{equation}
which in the $n\to\infty$ limit defines the functional integral 
measure with weight given by Eq. (\ref{e:libero}).

The  probability distributions $P_n$  define a Gaussian but not
stationary stochastic process. The correlation functions are readily
computed: $\langle Y(t) \rangle=0$ whereas if $t_1\le t_2$ the
autocorrelation is
$$
\langle Y(t_1) Y(t_2)\rangle= \frac{1}{\beta}\frac{t_1(L-t_2)}{L}
\qquad t_1\le t_2.
$$
It should be noted that the autocorrelation does not depend on
$|t_1-t_2|$ alone since the process is not stationary. We also observe
that due to the presence of $L$, which has the physical meaning of an
infrared cutoff (see below, Eq. (\ref{e:ris})) the process
(\ref{e:proc}) is not Markovian. Nevertheless for a set of $n$
successive times $t_1<t_2<\ldots t_n < L$ one verifies the property
$$
P_{1|n-1}(y_n,t_n|
y_1,t_1;y_2,t_2;\ldots;y_{n-1},t_{n-1})=P_{1|1}(y_n,t_n|y_{n-1},t_{n-1}),
$$
where $P_{l|k}(y_{k+1},t_{k+1};\ldots;y_{k+l},t_{k+l}|
y_1,t_1;y_2,t_2;\ldots;y_{k},t_{k})$ is the conditional probability of
observing $y_{k+1}$ at the time $t_{k+1}$ etc. given the $k$
events $y_1$ at the time $t_1$ etc. Our process is simply
related with the Wiener--L\'evy process \cite{kam}, defined by 
\begin{eqnarray}\label{e:wie}
&&P_1(y_1,0) = \delta(y_1) \nonumber \\
&&P_{1|1}(y_2,t_2|y_1,t_1) =
\left(\frac{\beta}{2\pi(t_2-t_1)}\right)^{1/2}\exp \left[ -
\frac{\beta}{2}\frac{(y_2 - y_1)^2}{(t_2-t_1)}
\right], 
\end{eqnarray} 
where $t_2>t_1$ and $-\infty < y < \infty$. It turns out that
(\ref{e:wie}) is simply the limit $L\to\infty$ of the process
(\ref{e:proc}).

Once we have  specified  unambiguously the stochastic process we are 
dealing with,  we turn to the calculation of  the mean energy of our
system. Given the boundary conditions $Y(0)=Y(L)=0$ we expand $Y$ in
Fourier modes as
$$
Y(t) = \sum_{n=1}^\infty A_n \sin \left( \frac{n\pi}{L}t\right). 
$$
The correlations for the coefficients $A_n$ are obtained from the
correlations for the process $Y(t)$. We get
\begin{eqnarray}\label{e:pl}
&&\langle A_n \rangle = 0 \nonumber\\
&&\langle A_n A_m \rangle = \delta_{n-m}\frac{2L}{\beta n^2\pi^2}.
\end{eqnarray}
In order to compute the mean energy we have to quantize the system:
the problem of the mean energy is of course the celebrated problem of
the black--body radiation solved by Planck.  We interpret the $A_n$ as
decoupled quantum harmonic oscillators.  We simply apply the Planck's
law changing in the l.h.s of Eq. (\ref{e:pl}) the frequency
distribution, i.e we perform the replacement:
$$
\frac{1}{\beta} \to  \frac{n\pi/L}{e^{\beta n\pi/L}-1}
$$
(we use units where $\hbar = 1$). Hence for the mean energy we have
\begin{eqnarray*}
\langle E \rangle && = \int {\mathcal D} y\, E[y] e^{-\beta E[y]} =
\frac{1}{4}\sum_{n=1}^\infty \frac{n^2\pi^2}{L^2} \,\langle A_n^2
\rangle \nonumber \\ && =\frac{1}{4}\sum_{n=1}^\infty
\frac{n^2\pi^2}{L^2}\frac{2L}{n^2\pi^2}\frac{n\pi/L}{e^{\beta
n\pi/L}-1} \to \frac{L}{2\pi}\int_0^\infty d\omega
\frac{\omega}{e^{\beta \omega} -1}.
\end{eqnarray*}
From this equation it follows immediately,
\begin{equation}\label{e:ris}
\langle E \rangle \equiv U = \frac{\pi L}{12}(k_B T)^2.
\end{equation}
Because, as we shall see in the following, the system defines a CFT,
from Eq. (\ref{e:ris}) we can read out the central charge, $c=1$ (see
for instance Ref. \cite{mal}). For the statistical entropy $S$ at the
temperature $T$ we find
$$
S = \frac{\pi L}{12}k_B^2 T.
$$
\subsection{The mapping between the Hilbert spaces}
Until now we have worked out the correspondence between 2D bosonic
string and stochastic process at the level of the spectra of the two
theories. In particular, it follows  that the two theories have the
same partition function. Let us now give simple arguments to show that
the correspondence holds also for the physical Hilbert spaces of the
two theories.  In a formal quantum treatment the 1D stochastic model
we considered is described by the means of an Hilbert space whose
structure is given by the familiar Fock space generated by creation
$\al_{-n}$ and annihilation $\al_n$ ($n>0$, $n\in {\Bbb N}$)
operators, with commutation relations $[\al_n,\al_m]=n\de_{m+n}$,
acting on a vacuum $\ket{0}$. But this space can be exactly mapped on
the Verma module (\ref{e:Verma}). Since the Verma module does not
contain null submodules, the one--to--one
correspondence between the respective states follows immediately.

\subsection{Conformal symmetry and stochastic processes}
Eq. (\ref{e:base}) gives the simplest realization of our idea. Of
course there are many other possibilities. An amusing example may be
the following. Let $I_\nu(z)$ be the Bessel function with imaginary
argument and $\mu = \sqrt{g^2 +1/4}$, where $g \in {\Bbb R}$. For
$y_0>0$ and $0<y_i<\infty$, $i=1,\ldots, n$; $y_{n+1}=y_0$, given the
time ordering (\ref{e:ord}) with $t_0=t_{n+1}=0$, we define the
process
\begin{eqnarray*}
&&P_n(y_1,t_1;y_2,t_2;\ldots;y_n,t_n) \nonumber \\ =
&&\left(e^{-\frac{\beta y_0^2}{L}}\,y_0 \,I_\mu(\frac{\beta
y_0^2}{L})\right)^{-1}
\frac{L}{\beta}\frac{\beta}{t_1}\frac{\beta}{t_2-t_1}\ldots
\frac{\beta}{L-t_n}\nonumber \\ && \cdot \prod_{j=1}^{n+1}
\left[\sqrt{y_jy_{j-1}} I_\mu\left(\frac{\beta
y_jy_{j-1}}{t_j-t_{j-1}}\right)\right]^{1/2}\exp \left[
-\frac{\beta}{2} \frac{(y_{n+1}-y_{n})^2}{t_{n+1}-t_{n}}\right].
\end{eqnarray*} 
If for a different ordering the process is again defined by the symmetry
condition $(ii)$ of Eqs. (\ref{e:Kolmo}), then all the Kolmogorov
axioms are obeyed. Again we have a non stationary and non Markovian
stochastic process, defined is such a way that (see Ref. \cite{kan}),
considering the limit $n\to\infty$, we get a functional integral 
measure with weight $e^{-\beta E[y]}$, where now $E[y]$ 
is given  by
$$
 E[y] = \frac{1}{2} \int_0^L dt\,
\left[\left(\frac{dy}{dt}\right)^2+ \frac{g^2}{y^2}\right].
$$
This establishes a link between conformal quantum mechanics \cite{dff} 
and stochastic process.
It is curious to observe that we can define a stochastic process only
if in the functional integration we have a conformally symmetric
weight. If a dimensional parameter is present, say a mass, then
there is no way to satisfy the crucial Martingale property $(iii)$ of
the Kolmogorov axioms (\ref{e:Kolmo}). 

\section{ Summary and Outlook}
In this paper we have shown that the map between string states and
local operators defined by the vertex operators in bosonic opens
string theory, can be used to put in correspondence, holographically,
open string theory with stochastic processes in one dimension.

The model we have described in this paper is too simple to be
considered more than a toy model.  However, it can be used to test
general ideas.  In particular it sheds light on the physical
meaning of the AdS/CFT correspondence in two dimensions, or more in
general on the holographic principle. The large $N$ limit of a
mechanical system in one spatial dimension, with short distance
behavior regularized by quantum mechanics, produces a field theory in
1+1 dimensions. Hence this mechanism can be used to generate one more
dimension and a field theory out of an ensemble of particles. Here
this is achieved in a non trivial way because we are going from a
simple mechanical system to a two-dimensional local field theory (the
bosonic open string), with a rich Lorentz and gauge structure.

Presently we do not know whether this mechanism is relevant also for
the AdS/CFT correspondence and the holographic principle in $d>2$.
Intuitively, it seems to be a basic and universal mechanism that could
work in general cases.


\begin{references}
\bibitem{adscft} J.~M.~Maldacena, {\em The Large $N$ Limit of
Superconformal Field Theories and Supergravity},
Adv.~Theor.~Math.~Phys.~{\bf 2} (1998) 231; Int.~J.~Theor.~Phys.~{\bf
38} (1999) 1113; \\ E.~Witten, {\em Anti-de~Sitter Space and
Holography}, Adv.~Theor.~Math.~Phys.~{\bf 2} (1998) 253; \\
S.~S.~Gubser, I.~R.~Klebanov and A.~M.~Polyakov, {\em Gauge Theory
Correlators from Non-Critical String Theory}, Phys.~Lett.~{\bf B428}
(1998) 105; \\ O.~Aharony, S.~S.~Gubser, J.~M.~Maldacena, H.~Ooguri
and Y.~Oz, {\em Large $N$ Field Theories, String Theory and Gravity},
Phys.~Rept.~323 (2000) 183.
\bibitem{adsdue}
A.\ Strominger, \JHEP  {\bf 01} (1999) 007;
M.\ Cadoni and S.\ Mignemi, \PRD {\bf D59} (1999) 081501; 
\NPB {\bf B557} (1999) 165;\PLB {\bf B490}, 131 (2000);
S.\ Cacciatori, D.\ Klemm and D.\ Zanon, [{\tt hep-th/9910065}];
S.\ Cacciatori, D.\ Klemm, W.A. Sabra and D.\ Zanon,
[{\tt hep-th/0004077}].
\bibitem{dff} V.~De Alfaro, S.~Fubini, and G.~Furlan,
{\em Conformal Invariance in Quantum Mechanics}, Nuovo Cimento {\bf 34A}
(1976) 569.
\bibitem{CCKM} M.~Cadoni, P.~Carta, D.~Klemm and S.~Mignemi,
{\em  AdS$_{2}$ gravity as conformally invariant mechanical system}, 
[\texttt{hep-th/0009185}].
\bibitem{CCK} M.~Cadoni, P.~Carta and  D.~Klemm ,
{\em  Large {N} limit of Calogero--Moser models and Conformal Field Theories}, 
in preparation.
\bibitem{ito}K. Ito, H. McKean, \textit{Diffusion processes and their
sample paths} (Springer Verlag, New York, 1965); G. Gallavotti,
\textit{Statistichal Mechanics. A short treatise} (Springer Verlag,
New York, 1999).
\bibitem {pol} J.\ Polchinski, {\it String Theory} (Cambridge Univ.\ Press,
Cambridge UK, 1998).
\bibitem{GSW} M.B. Green, J.H. Schwarz and E. Witten,
\textit{Superstring theory} (Cambridge Univ. Press, Cambridge UK, 1987). 
\bibitem{CC} M.~Cadoni and M.~Cavagli\`{a},
{\em Two-Dimensional Black Holes as Open Strings: A New Realization of the
AdS/CFT Duality}, [\texttt{hep-th/0005179}];{ \em Open Strings, 
2D Gravity and AdS/CFT Correspondence} [\texttt{hep-th/0008084}].
\bibitem{witten}
E.~Witten, Adv.~Theor.~Math.~Phys.~{\bf 2} (1998) 253;
\bibitem{br} R.C. Brower, Phys. Rev. D {\bf 6}, 1655 (1972);
R.C. Brower and P. Goddard, Nucl. Phys. B {\bf 40}, 437 (1972).
\bibitem{DDF} E. Del Giudice and  P. Di Vecchia, Nuovo Cim. {\bf 5A},
90 (1971); E. Del Giudice, P. Di Vecchia and S. Fubini,
Ann. Phys. {\bf 70}, 378 (1972).
\bibitem{dif} P. Di Francesco, P. Mathieu and D. S\'en\'echal
\textit{Conformal field theory} (Springer Verlag, New York, 1996).
\bibitem{gb} G. W. Gibbons and P.K. Townsend, \PLB {\bf B454} (1999) 187.
\bibitem{kam}N.G. Van Kampen, \textit{Stochastic processes in Physics
and Chemistry} (North Holland, Amsterdam, 1992).
\bibitem{mal} J. Maldacena and M. Strominger, \PRD  {\bf D56} (1997) 4975.
\bibitem{kan} S.F. Edwards and Y.V. Gulyaev, Proc. R. Soc. {\bf A279},
229 (1964); D.C. Khandekar and S.V. Lawande, J.Phys. {\bf A5}, 812
(1972). 


\end{references}
\end{document}